\documentclass[a4paper,11pt]{article}
\pdfoutput=1 

\usepackage{jheppub} 

\usepackage[T1]{fontenc} 
\usepackage[english]{babel}
\usepackage{ulem}
\usepackage{xcolor}
\usepackage{amsmath}
\usepackage{tensor}
\usepackage[utf8]{inputenc}
\usepackage{cleveref}

\title{Consistent Searches for SMEFT Effects in Non-Resonant Dilepton Events}
\preprint{\begin{flushright}MITP/18-124\\ZU-TH 47/18\end{flushright}}
\author{Stefan Alte${}^a$,}
\author{Matthias K\"onig${}^b$,}
\author[a,c]{and William Shepherd}

\emailAdd{stalte@uni-mainz.de}
\emailAdd{matthias.koenig@uzh.ch}
\emailAdd{shepherd@shsu.edu}

\affiliation[a]{PRISMA Cluster of Excellence \& Mainz Institute of Theoretical Physics, \\ Johannes Gutenberg-Universit\"at Mainz, 55099 Mainz, Germany}
\affiliation[b]{Physik-Institut, Universit\"at Z\"urich, CH-8057, Switzerland}
\affiliation[c]{Physics Department, Sam Houston State University, Huntsville TX 77341, USA}

\abstract{
Employing the framework of the Standard Model Effective Field Theory, we perform a detailed reinterpretation of measurements of the Weinberg angle in dilepton production as a search for new-physics effects. We truncate our signal prediction at order $1/\Lambda^2$, where $\Lambda$ denotes the new-physics mass scale, and introduce a theory error to account for unknown contributions of order $1/\Lambda^4$. Two linear combinations of four-fermion operators with distinct angular behavior contribute to dilepton production with growing impact at high energies. We define suitable angular observables and derive bounds on those two linear combinations using data from the Tevatron and the LHC. We find that the current data is able to constrain interesting regions of parameter space, with important contributions at lower cutoff scales from the Tevatron, and that the future LHC data will eventually be able to simultaneously constrain both independent linear combinations which contribute to dilepton production.
}

\begin{document}

\maketitle

\section{Introduction}

The Large Hadron Collider (LHC) is performing beyond expectations, delivering large samples of data. Aside from some interesting anomalies in the flavor sector, however, the collected data is largely in agreement with the predictions of the Standard Model (SM). In anticipation of (but without evidence for) new physics (NP), it is useful to constrain possible new effects in a framework as model-independent as possible.

One such framework is the Standard Model Effective Field Theory (SMEFT), in which a basis of local operators is built from the SM degrees of freedom (for a recent review, see~\cite{Brivio:2017vri}). In order to derive a complete and consistent operator basis, all operators allowed by the symmetries of the SM are considered and then reduced using equations of motion of the degrees of freedom of the theory, integration by parts identities, and Fierz transformations~\cite{Grzadkowski:2010es}. It is crucial to keep in mind that this utilizes field redefinitions which are only defined up to the order of the operator product expansion, meaning that the fields in a dimension-six Lagrangian are only defined up to $\mathcal O (1/\Lambda^2)$ in the power-counting of the effective theory. This has nontrivial implications for the interpretation of calculations in the SMEFT~\cite{Barzinji:2018xvu,Helset:2018dht,Criado:2018sdb}.

Two particularly important consequences arise from this: First, when one computes amplitudes and scattering cross sections in this framework, a consistent power-counting scheme should be followed. When squaring an amplitude containing contributions at $\mathcal O (1/\Lambda^2)$ and lower, the $\mathcal O (1/\Lambda^4)$ piece is of the same order as an interference term between the $\mathcal O (\Lambda^0)$ and any uncalculated $\mathcal O (1/\Lambda^4)$ contribution. Without an operator basis at dimension eight for the higher-dimensional contribution, it is not possible to calculate the fulll term of $\mathcal O(1/\Lambda^4)$, and it should thus be treated as an uncertainty. This is in analogy to e.g. QED calculations, where the one-loop calculation only predicts the cross-section with corrections of $\mathcal O(\alpha)$; the squared loop diagrams cannot be included in the prediction until the full two-loop analysis has been performed. If they are improperly included regardless, one finds that the contribution of those graphs at $O(\alpha^2)$ is infinite, which indicates the inconsistency of the calculation.

The second consequence is the fact that the higher-dimensional operators induce shifts in the electroweak input parameters: If effects from these operators are present, they influence the extraction of electroweak parameters from data, resulting in corrections starting at $\mathcal{O}(v^2/\Lambda^2)$, which apply to the result of numerous would-be purely SM calculations, and also can be used to constrain the SMEFT Wilson coefficients in conjunction with electroweak precision data~\cite{Berthier:2015oma,Berthier:2015gja,Bjorn:2016zlr,Berthier:2016tkq,Ellis:2018gqa,Almeida:2018cld}. The effect of these shifts in the SM parameters can be disentangled to some extent from the more direct contributions because of their different behavior with event energy; the direct contributions generically grow in importance relative to the SM process by a factor of ${s}/{\Lambda^2}$, while the shift-type contributions do not. In this article we focus on the effects growing in importance with energy, and neglect the input parameter shift contributions, which are already well constrained by precision measurements at lower energy.

When using hadron collider data to constrain the effective couplings, the large momentum transfer in the partonic hard scattering can produce events close to the cutoff scale of the effective theory. The canonical treatment of these events has been to determine which of them exceed some threshold beyond which we expect the EFT is not reliable and discard those events~\cite{Englert:2014cva,Contino:2016jqw,Farina:2016rws}. As shown in~\cite{Alte:2017pme}, however, treating the squared dimension-six contribution as an uncertainty leads to these events being naturally discarded from the fit as the uncertainties grow faster with energy than the interference term between the SM and dimension-six amplitudes, while simultaneously estimating honestly the uncertainty in the region where it is neither dominant nor negligible. This prescription yields conservative bounds on the possible values of certain linear combinations of Wilson coefficients in the SMEFT; in conjunction with tools to expedite matching of UV theories onto the SMEFT~\cite{Criado:2017khh,Celis:2017hod,Bakshi:2018ics} this allows for constraints to be rapidly applied to newly-invented theories far into the future, making this a particularly useful form in which to cast LHC and other precision measurements to ensure future utility.

In this work, we study SMEFT effects in dilepton production, carefully taking into account both concerns of allowing multiple operators simultaneously and theoretical uncertainties due to the truncated perturbation expansion in $\Lambda^{-1}$. In~\cite{Alte:2017pme} this analysis was performed for SMEFT contributions to non-resonant dijet production at the LHC. The parameter space regions which were excluded there were characterized by two qualities: First, there is either a minimum value for the NP scale $\Lambda$ or a maximum value for the Wilson coefficients, beyond which the signal from NP would become large enough to conflict with the LHC measurements. However, regions of very large couplings or small NP scales are not excluded by the data, as the EFT errors become so sizable that it is impossible to predict what behavior is expected with enough accuracy to then constrain it. These features are generic to any consistent treatment of EFT errors, and will be present in this analysis as well.

In order to close these windows in parameter space at lower EFT cutoff scale or higher Wilson coefficient, data at lower energies, where the perturbation series is better behaved, is needed. For this reason, we include here a recast from dilepton production at the Tevatron in our analysis and show the complimentarity of the results. We note that in principle one could also include data from dijet production at LEP or dilepton production at the SPS to constrain the low-scale parameter space even further, but leave this for potential future work; it is in any case likely that the UV completions of SMEFT candidates with cutoff scales lower than those constrained by the Tevatron analysis would have been directly probed in later experiments.

In the next section, we review SMEFT contributions to dilepton production, with a particular focus on which operators have distinguishable behaviors at leading order in the power counting. We discuss the general strategy used to design searches for these four-fermion operators in \cref{sec:genstrat}, and then recast and present future reaches for LHC searches for these effects in \cref{sec:LHC}. In \cref{sec:TeV} we investigate the bounds which can be derived from the Tevatron legacy measurement of the Weinberg angle, and we conclude in \cref{sec:conc}.

\section{Dilepton Production in the SMEFT}
\label{sec:SMEFTeffect}

In the SMEFT the SM is extended by local, higher-dimensional operators built from SM fields. These higher-dimensional operators are invariant under the SM gauge group $SU(3)_C \times SU(2)_L \times U(1)_Y$. The SMEFT Lagrangian can be written as
\begin{equation}
\label{eq:lagrangian}
 \mathcal{L}_\text{SMEFT} = \mathcal{L}_\text{SM} + \mathcal{L}^{(5)} + \mathcal{L}^{(6)}+\mathcal{L}^{(7)}+\mathcal{L}^{(8)} + \dots \,,
\end{equation}
where $\mathcal{L}_\text{SM}$ has the form of the SM Lagrangian but the couplings receive corrections scaling as $v^2/\Lambda^2$ and $\mathcal{L}^{(i)}$ with $4 < i$ denotes the Lagrangian contributions from operators of mass dimension $i$. The Lagrangian $\mathcal{L}^{(i)}$ exhibits the form
\begin{equation}
 \mathcal{L}^{(i)} = \sum_{k=1}^{N_i} \frac{C_k^{(i)}}{\Lambda^{i-4}} Q_k^{(i)},
\end{equation}
where $C$ are the Wilson coefficients, $Q$ are the operators and $\Lambda$ is the NP scale. The sum includes all the $N_i$ non-redundant operators at the corresponding mass dimension. Operator bases up to dimension eight are known~\cite{Weinberg:1979sa,Wilczek:1979hc,Buchmuller:1985jz,Grzadkowski:2010es,Abbott:1980zj,Lehman:2014jma,Lehman:2015coa,Liao:2016hru}. At the level of dimension five only the Majorana mass operator~\cite{Weinberg:1979sa,Wilczek:1979hc} exists. We work at the dimension-six level, where the leading contribution to dilepton production arises, and employ the basis commonly referred to as the ``Warsaw basis''~\cite{Grzadkowski:2010es}.

The baryon-number and CP-conserving four-fermion operators contributing at highest order in center-of-mass energy to dilepton production in the limit of $SU(3)^5$ flavor symmetry at dimension six are shown in \cref{tab:dilepton_operators}, where we employ the conventions from~\cite{Brivio:2017btx}.
\begin{table}
\renewcommand{\arraystretch}{1.4}
\centering
\begin{tabular}{rc|ccrc|c}
\cline{1-3} \cline{6-7} 
&$Q^{(1)}_{lq}$ & $ \left( \bar{l}_p \gamma_\mu l_p \right) \left( \bar{q}_s \gamma^\mu q_s \right)$ & & 
&$Q_{lu}$ & $ \left( \bar{l}_p \gamma_\mu l_p \right) \left( \bar{u}_s \gamma^\mu u_s \right)$  \\
&$Q^{(3)}_{lq}$ & $\left( \bar{l}_p \gamma_\mu \tau^I l_p \right) \left( \bar{q}_s \gamma^\mu \tau^I q_s \right)$ & &
&$Q_{ld}$  & $ \left( \bar{l}_p \gamma_\mu l_p \right) \left( \bar{d}_s \gamma^\mu d_s \right)$ \\
&$Q_{eu}$ & $ \left( \bar{e}_p \gamma_\mu e_p \right) \left( \bar{u}_s \gamma^\mu u_s \right)$ & &
&$Q_{qe}$  & $ \left( \bar{q}_p \gamma_\mu q_p \right) \left( \bar{e}_s \gamma^\mu e_s \right)$  \\
&$Q_{ed}$ & $ \left( \bar{e}_p \gamma_\mu e_p \right) \left( \bar{d}_s \gamma^\mu d_s \right)$ & &
\end{tabular}
\caption{The baryon-number and CP-conserving operators contributing to dilepton production in the flavor-symmetric limit at dimension six in the Warsaw basis~\cite{Grzadkowski:2010es} as four-fermion contact interactions.}
\label{tab:dilepton_operators}
\end{table}
We denote the left-handed quark and lepton doublets by $q$ and $l$. The right-handed up-quark, down-quark and lepton singlets are $u$, $d$ and $e$. The Pauli matrices are $\tau^I$, where $I\in{1,2,3}$, and $p,s$ are generation indices for SM fermions. We focus in this work on the operators whose contribution dominates in the limit of high energies; additional operators contribute (approximately) as a linear shift of the SM amplitude due to corrections of the couplings in the SM; these shifts have been fairly well studied in the context of LEP and other precision experiments~\cite{Berthier:2015oma,Berthier:2015gja,Bjorn:2016zlr,Berthier:2016tkq,Ellis:2018gqa,Almeida:2018cld}, and are in any case subdominant effects at higher energies, so we neglect them here.

The dilepton production cross section, including the effects of dimension six operators, can be written as
\begin{align}
 \sigma                    &= \sigma_\text{SM} + \frac{1}{\Lambda^2} \, \sigma_\text{int} + \frac{1}{\Lambda^4} \sigma_\text{BSM} + \dots \,,
 \label{eq:dilepton_cs}
\end{align}
where $\sigma_\text{SM}$ is the contribution from the SM operators, $({1}/{\Lambda^2}) \, \sigma_\text{int}$ is the contribution from the interference of the dimension-six operators with the SM piece and $({1}/{\Lambda^4}) \, \sigma_\text{BSM}$ denotes the contribution from the squared dimension-six piece. Truncating the series in inverse powers of the NP scale as in \cref{eq:dilepton_cs} is inconsistent, as the contribution from the interference of unspecified dimension-eight operators with the SM contributes at the same order in ${1}/{\Lambda}$ as $\sigma_{\text{BSM}}$. There is debate in the community regarding whether or not $\sigma_{\text{BSM}}$, the squared dimension-six amplitude, is well-defined under change of basis, as well. Even if it is ultimately shown to be well-defined, having a signal function linear in the Wilson coefficients significantly increases the ease of implementing global fits of the SMEFT utilizing multiple different observables. Thus, we truncate our signal after the second term in \cref{eq:dilepton_cs},
\begin{align}
\left.\sigma\right|_\text{signal} &=  \sigma_\text{SM} + \frac{1}{\Lambda^2} \, \sigma_\text{int}  \,,
\end{align}
and use the dimension-six squared piece as an ansatz for the uncertainty from higher orders in the SMEFT power-counting. We note that it is possible that this formula yields a negative cross section for large enough Wilson coefficients or small enough NP scales. However, in all regions where this is the case the next-order term which we adopt as a theoretical error is larger still, effectively removing all bins with questionable predictions from a statistical analysis.

The effects of some four-fermion operators on dilepton production at the LHC can be distinguished from each other based on their angular behavior. Each of the relevant operators in \cref{tab:dilepton_operators} contributes to one of two distinguishable angular distributions with some strength, leading us to define two linear combinations of Wilson coefficients which parameterize the total effect of all these operators. At the partonic level, this can be understood as each operator contributing dominantly to either forward or backward production of the lepton, as defined in the center of mass frame where the direction of travel of the initial-state quark is defined to be forward. We thus will label these linear combinations as $c_{\rm fwd}$ and $c_{\rm bwd}$.

At a hadron collider, we unfortunately do not have unambiguous definitions of forward or backward available to us, as it is possible to have selected a quark rather than an antiquark out of either initial-state hadron. If we consider the case of the Tevatron, it's reasonable to expect that a quark more likely came out of the proton than the antiproton, and thus we define forward relative to the proton's direction of travel. While this will not be accurate for every event, on average it will still be more often correct.

At the LHC, where both beams are protons and thus there is no a priori good proxy for forward, we instead must depend on the fact that valence quarks tend to carry relatively large fractions of the proton momentum relative to the sea quarks, which are just as likely to be a quark as an antiquark. As a result, the overall boost of the collision system is more often in the direction the quark was traveling. Therefore, interactions which tend to produce a forward lepton in the partonic center of mass frame generally give a higher absolute value of pseudo-rapidity $\eta$ to the lepton in the lab frame. We therefore define the events which are ``forward'' to be those where the lepton has a greater $|\eta|$ than the antilepton.

The linear combinations which describe SMEFT effects on dilepton production are given by
\begin{equation}
\begin{aligned}
\label{eq:lincombs}
c_{\rm fwd}&=C^{(3)}_{lq} - 0.48\, C_{eu} - 0.33\, C^{(1)}_{lq}+ 0.15\, C_{ed}\,, \\
\text{and}\quad c_{\rm bwd}&=C_{lu}+0.81\, C_{qe}-0.33\, C_{ld}\,.
\end{aligned}
\end{equation}

The precise coefficients describing the impact of each operator on the overall signal rate are dependent on PDF effects, and thus run slightly as the center of mass energy changes; we anticipate that any measurement based on these effects will be much more challenging than the overall determination of the strength of signal due to a particular linear combination, and thus do not consider this possibility further here.

\section{General Strategy for the Searches}
\label{sec:genstrat}

As discussed in the previous section, the different operators entering the NP-contribution to dilepton production can be disentangled to some extent by studying their angular spectra. Since the combinations in \cref{eq:lincombs} prefer either forward or backward events, we perform the search simultaneously in the total rate as well as the asymmetry between the forward and backward events.\footnote{Note that current LHC and Tevatron measurements do not report the  asymmetry for the full range of dilepton invariant masses; for search recasts using current data we thus only consider the total rate.} By doing so, we recover some of the constraining power of the fit even in cases where the effects from $c_\mathrm{fwd}$ and $c_\mathrm{bwd}$ cancel in the total rate but not the angular spectrum.

The asymmetry and the rate are defined as
\begin{align}
 A_{FB} = \frac{N_F-N_B}{N_F+N_B}\,, \qquad\qquad N_\mathrm{tot} = N_F + N_B \,,
\end{align}
where $N_F$ and $N_B$ refers to the number of events where the negatively charged lepton is either more forward or more backward with respect to the positively charged one. As motivated in section~\ref{sec:SMEFTeffect} we expand both quantities in inverse powers of the NP scale. As discussed, our signal contains only the pieces up to order ${1}/{\Lambda^2}$, and we implement a theoretical error which corresponds to neglected contributions at order $1/\Lambda^4$. Our total error is determined by adding the statistical error, the systematic error and the theory error in quadrature. We employ a Poisson error for the statistics and comment on our treatment of the systematics in the specific searches in sections~\ref{sec:LHC} and~\ref{sec:TeV}. Our theory error is implemented as follows: we combine the squared dimension-six piece and the two pieces corresponding to dimension-eight interference in the forward and backward bin in quadrature. While we explicitly use our Monte-Carlo data for the squared piece, we model the dimension-eight interference by symmetrizing the dimension-six piece and substituting the squared Wilson coefficients according to 
\begin{align}
 C_k^2 \rightarrow g_W^2 \, C_8 \, \sqrt{N_8} \,,
\end{align}
where $g_W$ is the $SU\left(2\right)_L$ gauge coupling, $C_8=\text{max}\left(1,|C_k|\right)$ is a proxy for the expected coupling of an unknown dimension-eight operator, and $N_8$ is a guess for the number of dimension-eight operators expected to independently contribute, meaning that larger values of $N_8$ correspond to a more conservative error treatment and hence weaker bounds. We chose illustrative values for $N_8$ which yield results allowing the understanding of the impact of this parameter on the resulting bounds. Given that seven operators contribute at dimension-six and there are approximately 10 times as many dimension-eight as dimension-six operators~\cite{Henning:2015alf}, a reasonable guess might be $N_8=70$, though it is not clear that all those operators would lead to maximal energy growth.\footnote{The development of a full basis of dimension-six operators would enable precise determinations of $N_8$, and also allow for analyses that are complete at order $\Lambda^{-4}$. It would also, however, introduce a large number of new parameters requiring constraint. Since the SMEFT at order $\Lambda^{-2}$ is currently underconstrained, we do not see this as a fruitful path forward at this time.} We choose values for our figures which best illustrate the impact of this error parameter on the analysis and resulting bounds. In this procedure we allow for our ignorance of the structure of contributions at dimension eight by allowing for independent fluctuations in the forward and backward bins of the cross section. In our analysis, we switch on only the two Wilson coefficients $C^{(3)}_{lq}$ and $C_{lu}$ as proxies for the corresponding linear combinations from eq.~\eqref{eq:lincombs}.

Other schemes of ensuring the validity of the EFT expansion have been explored in the literature. An intuitive example of such a scheme is to discard events above a certain energy scale that one associates with the cutoff of the effective theory. The key difference to our approach is the fact that this represents a hard cutoff, equivalent to an assumed infinite-magnitude step function as the estimated theory error. In this scheme, events very close to the cutoff are treated as having no uncertainty from the EFT treatment when they are in fact relatively unreliable, but not completely unreliable. Our method instead acts as a smooth cutoff, de-weighting events as they get closer to the edge of EFT validity.

We use partonic Monte-Carlo pseudodata generated at LO with \texttt{MadGraph5}~\cite{Alwall:2014hca}; it is well known that observables with light leptonic final states are not strongly affected by parton showering and detector simulation effects. Both the renormalization and factorization scales are chosen dynamically to be the sum of the transverse energies of the produced leptons. For the search at the Tevatron, we employ an identification efficiency correction, further specified below. We use the implementation of the SMEFT operators from the \texttt{SMEFTsim} package~\cite{Brivio:2017btx,Aebischer:2017ugx}, and employ a $\chi^2$-test to derive our bounds at the 95\% confidence level, treating the total error in each bin as described above.

\section{Searches at the LHC}
\label{sec:LHC}
We model our proposed search on the results reported by ATLAS in~\cite{Aaboud:2017buh} corresponding to $36.1\text{ fb}^{-1}$ of data taken at a center-of-mass energy of $13\text{ TeV}$. Our search uses seven bins in the dilepton invariant mass ranging from $400\text{ GeV}$ to $6000\text{ GeV}$. As a validation, we compare our Monte-Carlo results for the SM Drell-Yan total cross section with the Drell-Yan background predictions from table~3 in~\cite{Aaboud:2017buh}, with which they agree well. This search requires that both the electron and the positron pseudorapidity $\eta$ lie in one of the ranges $2.47>|\eta|>1.52$ or $1.37 > |\eta|$, to ensure accurate tracking and calorimetric information, and that the transverse momenta of the electron and the positron fulfill  $p_T > 30~\mathrm{GeV}$ . We use the CT10 PDF~\cite{Lai:2010vv} throughout the LHC analysis. We assume that the asymmetry is measured with negligible systematic uncertainty and use the systematic errors quoted by ATLAS for the total rate. In our fits at higher integrated luminosities, we assume that these systematics fall off with the square root of the luminosity to model improved control-region statistics, but do not fall below 2 \% of the event number.

\begin{figure}
\centering
\includegraphics[scale=0.55]{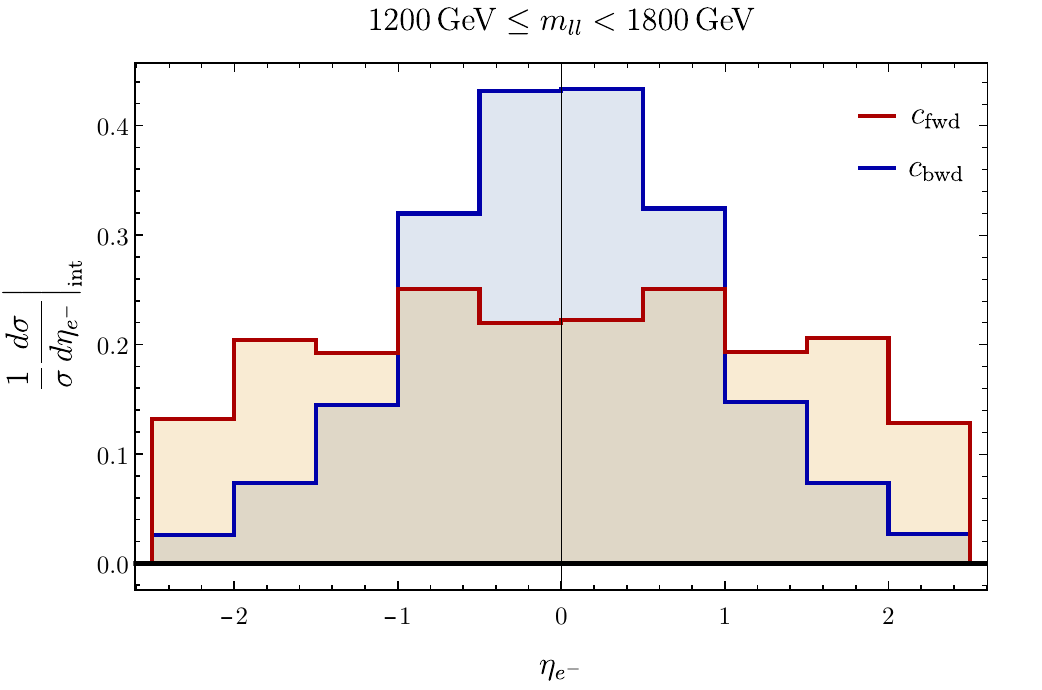}
\caption{Normalized interference cross-sections at the LHC differential in the pseudorapidity $\eta$ of the electron. The red spectrum shows the contribution generated by the operators of the combination $c_\mathrm{fwd}$ in \cref{eq:lincombs}, while the blue spectrum shows the contribution from $c_\mathrm{bwd}$.}
\label{fig:eta_spec_LHC}
\end{figure}

Binning the interference term in the pseudorapidity of the electron in ten bins of width 0.5 between -2.5 and 2.5 we find the spectra shown in \cref{fig:eta_spec_LHC}. The plots are grouped by linear combination. It is clear here that the ``forward'' class of operators tends to give a larger $|\eta|$ to the lepton than the ``backward'' class, as discussed in \cref{sec:SMEFTeffect}. Note that the apparent downward fluctuation of the bins where $1.0<|\eta|<1.5$ is due to the fact that a portion of this range is explicitly vetoed in the analysis.

This difference in angular behavior motivates a two-dimensional binning where one observable is the dilepton invariant mass. The other observable is defined by the pseudorapidities of the electron $\eta\left(e^-\right)$ and the positron $\eta\left(e^+\right)$ where the ``backward'' bin contains the events with $\left|\eta\left(e^-\right)\right|<\left|\eta\left(e^+\right)\right|$ and the ``forward'' bin contains the events with $\left|\eta\left(e^+\right)\right|<\left|\eta\left(e^-\right)\right|$. We note that this asymmetry is not reported in~\cite{Aaboud:2017buh}, and thus only employ the asymmetry variable for projected future searches. We strongly advocate that future measurements of this process report this important angular data in addition to the cross section differential in the invariant mass to enable future constraints based fully on experimental data.

\begin{figure}
\centering
\includegraphics[scale=0.5]{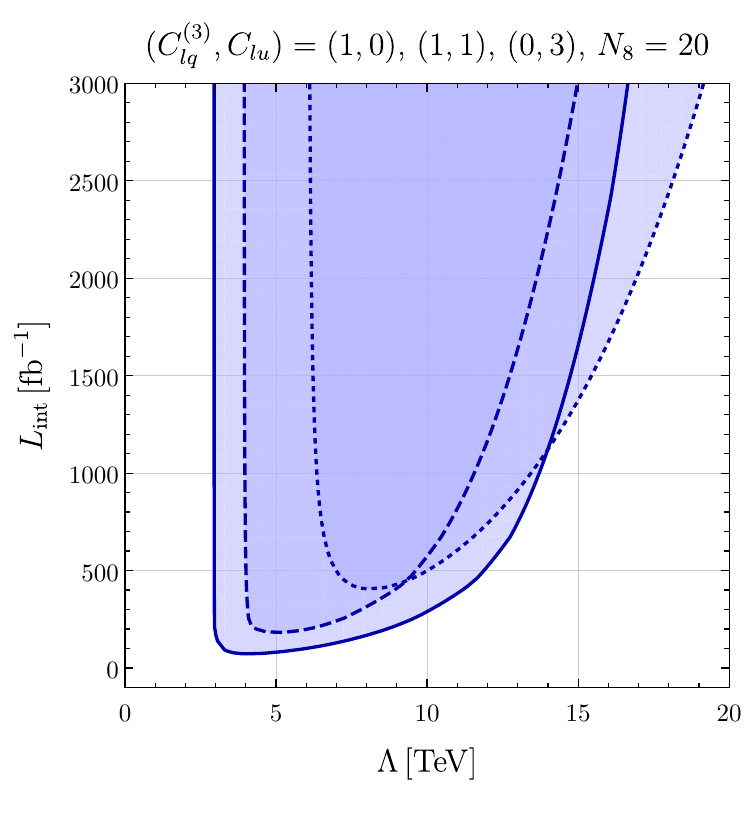}
\includegraphics[scale=0.5]{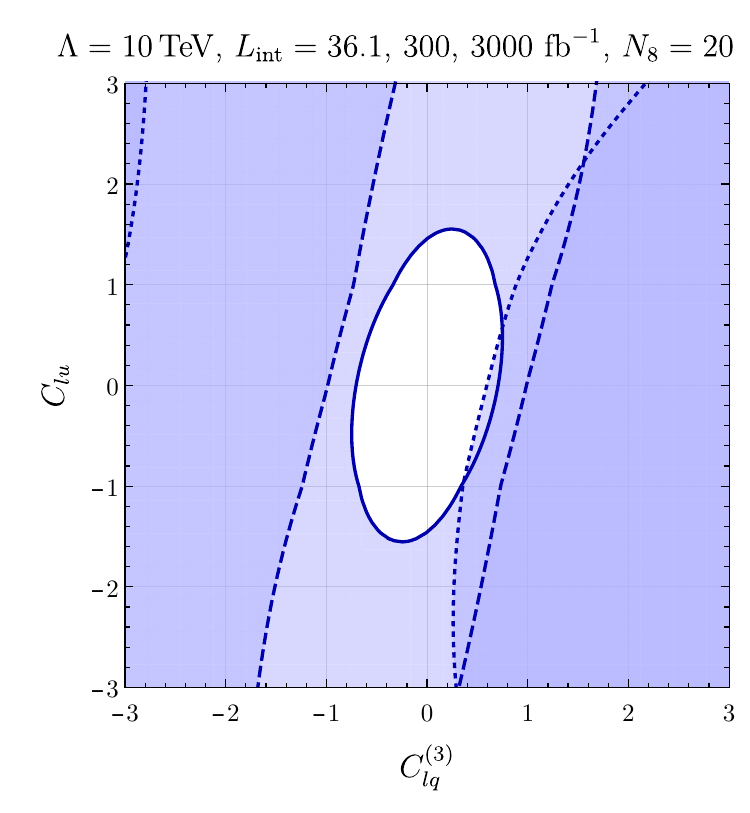}
\caption{Exclusion regions from proposed LHC searches. In the left panel, we show exclusion regions for fixed values of Wilson coefficients as functions of the scale $\Lambda$ and the integrated luminosity $L_\mathrm{int}$. Solid, dashed and dotted lines correspond to the combinations $(C_{lq}^{(3)},C_{lu})=(1,0)$, $(1,1)$ and $(0,3)$, respectively. In the right panel, the scale $\Lambda$ is fixed and the exclusion regions in the plane of the two Wilson coefficients is shown for $L_\mathrm{int}=36.1, 300$ and $3000~\mathrm{fb}^{-1}$.}
\label{fig:excl_asym_LHC}
\end{figure}

Our exclusion plots are shown in \cref{fig:excl_asym_LHC}. The constraints arising from the currently public LHC analysis, which does not report the asymmetry necessary to successfully differentiate between the two linear combinations of Wilson coefficients, are mainly able to constrain coupling combinations which increase the total dilepton rate at high energies. This is because there is a slight deficit in the high-$m_{ll}$ tail relative to the predicted SM background. Note that the current constraints are distorted by the deficit in data in~\cite{Aaboud:2017buh} relative to the SM prediction, and thus are not directly comparable to the future reach predictions, which assume that the data will agree with the SM prediction. The minimum coupling constraints from future searches with large amounts of data roughly form ellipses with their major axes oriented along the direction of cancellation between the two linear combinations in the total cross section of dilepton production; the  asymmetry measurement alone is what ultimately constrains larger excursions in this direction in parameter space. Note that the curve for $300~{\rm fb}^{-1}$ does not quite close in this direction; the earliest closure we find for $N_8=20$ occurs for approximately 400 fb$^{-1}$ of data. Considering the more pessimistic choice of $N_8=70$, the curves close only after approximately $750~{\rm fb}^{-1}$ are analyzed, and the ultimate sensitivity expands from a semi-major axis length of $\sim1.5$ (as seen in \cref{fig:excl_asym_LHC}) to one of $\sim2.5$.

We find that the constraints from the LHC for $\Lambda=10$ TeV extend to quite high values of the Wilson coefficients, with the smallest maximum coefficient constrained using $L_{\rm int}=3000$ fb$^{-1}$ at the point $\left(C_{lq}^{(3)},C_{lu}\right)\sim\left(\pm2,\pm11\right)$. These Wilson coefficients are at least nearing non-perturbativity, so the inability to reliably calculate the signal prediction at points beyond this value perturbatively is not startling.

\section{Searches at the Tevatron}
\label{sec:TeV}
We model our search at the Tevatron on the recent measurement of $\theta_W$ by the CDF collaboration~\cite{Aaltonen:2016nuy}. This exploits $9.4\text{ fb}^{-1}$ of data taken at a center-of-mass energy of $1.96\text{ TeV}$. We use the CTEQ5L PDF~\cite{Lai:1999wy}, in keeping with the CDF approach to pseudodata generation. We follow the analysis in distinguishing two different acceptance regions referred to as ``central-central'' (CC) and ``central-plug'' (CP); the ``plug-plug'' region was significantly less populated in this analysis and suffered from the highest uncertainties due to mismeasurement, so we neglect it in our analysis. The selection criteria in the CC region are: one of the transverse energies of the electron and the positron larger than $15\text{ GeV}$ and one transverse energy larger than $25\text{ GeV}$; both pseudorapidities have to fullfill $0.05< | \eta| <1.05$. The selection critera in the CP region are: both transverse energies larger than $20 \text{ GeV}$; one pseudorapidity with $0.05 < |\eta| < 1.05$ and the other pseudorapidity with $1.2 < |\eta|< 2.8$. We follow the CDF collaboration in using a different binning in the dilepton invariant mass in each topology. Combining multiple of their individual bins, we use bin borders of
\begin{equation}
\begin{aligned}
\text{CC:} \qquad m_{ll} &\in \{130,162,203,255,320,400 \} \text{ GeV} \,, \\
\text{CP:} \qquad m_{ll} &\in \{130,163,205,256,321,402 \} \text{ GeV}  \,.
\end{aligned}
\end{equation}
As a validation, we again compare our SM predictions in both regions with the data from Tevatron. After introducing one overall efficiency factor for each distinct event topology (CC and CP) and fitting them to the CDF backgrounds, our background predictions agree up to deviations negligible for our purposes. 

With an understanding of the CDF efficiencies and SM background, we then apply the same efficiencies to the signal distribution pseudodata and investigate what values of the Wilson coefficients can be constrained by the data studied by CDF. The CDF measurement reports $A_{FB}$ in the range $64\mathrm{\ GeV}<m_{ll}<150 \mathrm{\ GeV}$, as well as in underflow and overflow bins, but does not provide a breakdown of this asymmetry at higher invariant masses, which is necessary to get the full constraining power available from this data. We thus neglect the asymmetry observable in our recast of that data, but note that including that information in an updated measurement would strengthen our bounds somewhat.

We utilize the CDF data with errors due to statistics calculable from the number of events in each bin. The CDF collaboration has not separately reported systematic errors in this analysis, and they are not readily accessible from the plots due to their relative smallness as compared to the statistical errors. We therefore hypothesize an overall systematic error of $5\%$, though we have confirmed that the overall measurement remains very much statistics dominated under any reasonable assumption for this systematic error magnitude. 

\begin{figure}
\centering
\includegraphics[scale=0.5]{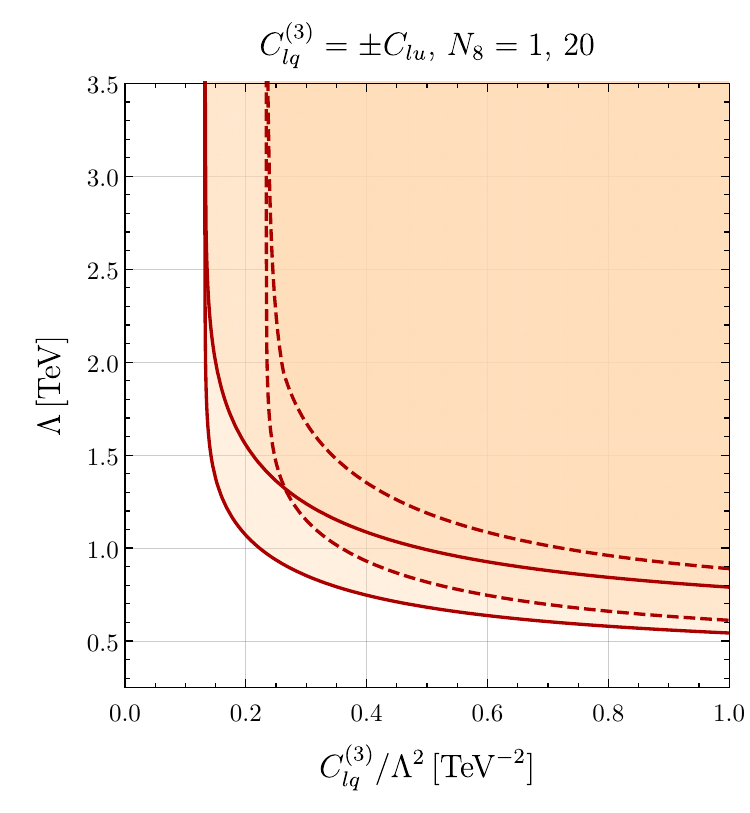}
\includegraphics[scale=0.5]{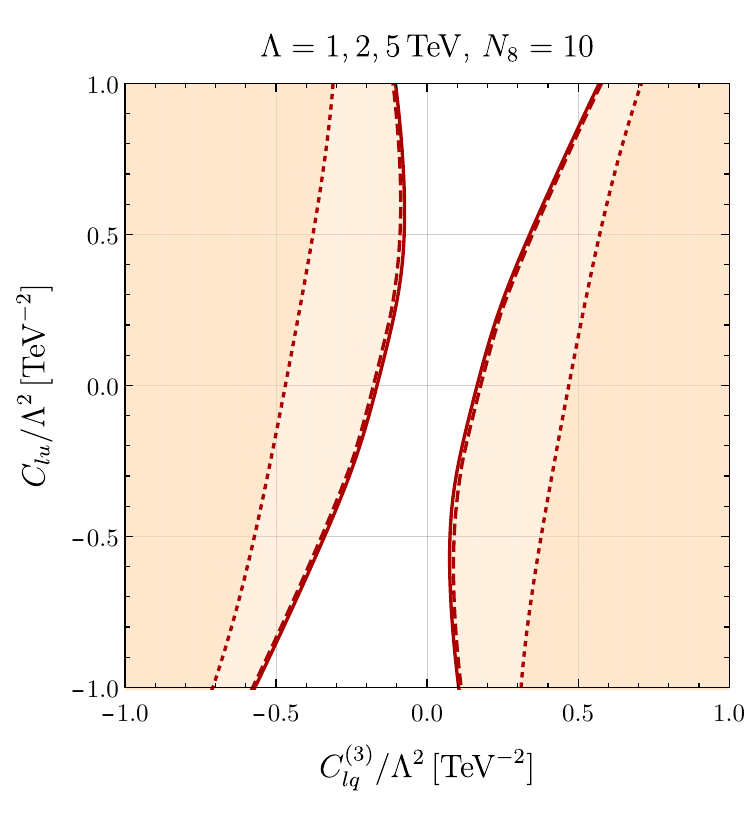}
\caption{Exclusion regions from the Tevatron data. In the left panel, we choose either $C_{lq}^{(3)}= -C_{lu}$ (solid lines) or $C_{lq}^{(3)}= +C_{lu}$ (dashed lines). The choice of the parameter $N_8$ affects the value of the scale $\Lambda$ at which constraints can be found: The more constraining regions (extending to lower values of $\Lambda$) belong to $N_8=1$ while the less constraining region corresponds to $N_8=20$. In the right panel, we show exclusions when the scale is fixed to $\Lambda= 1,2,5$~TeV (dotted, dashed and solid lines). Note that, in both figures, the Wilson coefficients have been normalized such that any difference between the curves for different scales $\Lambda$ is due to independent sensitivity of the data to the Wilson coefficient and the NP scale.}
\label{fig:excl_asym_Tev}
\end{figure}

We then find the excluded regions of this two-dimensional parameter space and present the constraints in \cref{fig:excl_asym_Tev}. We derive constraints for both cases of fixed-coefficient and fixed-scale. We note that, for scales above approximately 2.0 TeV for $N_8=1$ and 3.5 TeV for $N_8=20$, the bounds have the general dependence on ${C}/{\Lambda^2}$ that one would na\"\i vely suspect, but for lower scales the degeneracy between Wilson coefficient and NP scale is broken by the presence of large dimension-eight effects in the theory error. The statistics available at the Tevatron allow for the reliable measurement of only one of the two linear combinations of four-fermion operators once appropriate theory errors are included in the analysis; $C_{lq}^{(3)}$ is constrained, with the precise value of those constraints depending only weakly on $C_{lu}$, consistent with their relative impact on the total dilepton rate. We find that the breakdown of the EFT expansion prevents Tevatron data from constraining SMEFT effects due to $|C_{lq}^{(3)}|> 5.7\,{\Lambda^2}/{\mathrm{TeV^2}}$ for $C_{lu}=0$. Considering the more pessimistic assumption for theoretical errors of $N_8=70$, the curve for $\Lambda=1~{\rm TeV}$ is weakened by a linear shift of approximately $0.4~{\rm TeV}^{-2}$; the other curves move only very slightly.

\begin{figure}
 \centering
 \includegraphics[scale=0.5]{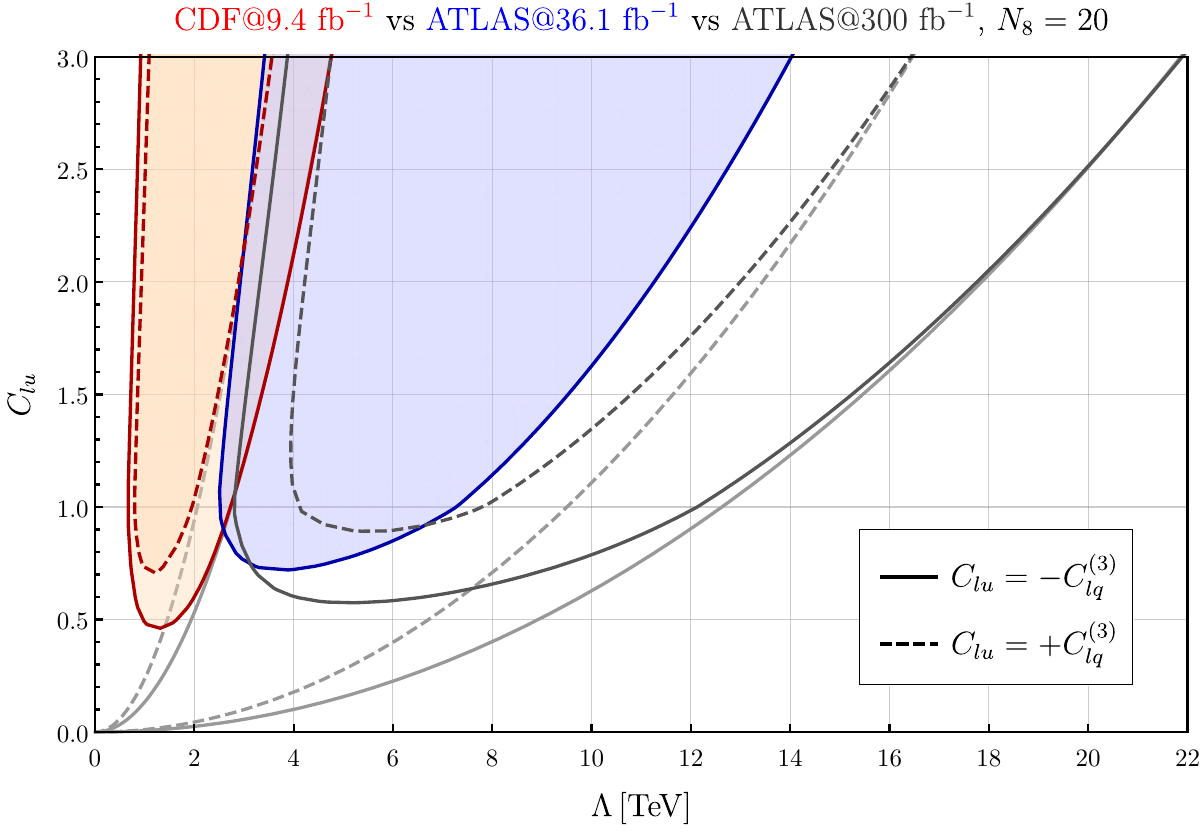}
 \caption{Combined exclusion regions from the CDF data (red) and the pseudodata for ATLAS at $36.1~\mathrm{fb}^{-1}$ (blue) and $300~\mathrm{fb}^{-1}$ (dark gray). It can be seen that the data from lower energies closes the region not covered by the ATLAS search since the EFT expansion is valid for lower values of $\Lambda$. The light gray curves approximate the result of the search for CDF and future ATLAS data if the limiting behavior of dependence only on the ratio ${C}/{\Lambda^2}$ were assumed throughout the parameter space; note that everywhere this fails the actual bounds are weaker.}
 \label{fig:excl_asym_comb}
\end{figure}

Despite the relatively limited statistics at the Tevatron, these bounds still fill an important role relative to the LHC constraints and future reach. The inability of the LHC to constrain effects at lower scales due to the fairly high characteristic energy of its collisions leaves a gap in its sensitivity below a few TeV which Tevatron data can at least partially fill. This is illustrated in \cref{fig:excl_asym_comb}, which compares the full Tevatron dataset constraints and the current LHC constraints on the same footing. It's manifest that the CDF measurement has constrained a region of parameter space which is inaccessible to the LHC due to the theoretical errors associated with the EFT expansion in the limit where the partonic collision energy has become comparable to or greater than the NP scale. In this figure it is again possible to see the separate sensitivity of each machine to the NP scale and the Wilson coefficient; if the na\"\i ve dependence on only ${C}/{\Lambda^2}$ held all of these constraints would be parabolae passing through the point $(0,0)$ in this figure, described by ${C}/{\Lambda^2} = {1}/{M^2}$. These curves are shown in solid gray lines for the case $C_{lq}^{(3)}=-C_{lu}$ for 300 fb$^{-1}$ of data at the LHC ($M=12.63$~TeV) and the current Tevatron data ($M=2.74$~TeV) and in dashed gray lines for  $C_{lq}^{(3)}=C_{lu}$ with $M=9.50$~TeV at the LHC and $M=2.06$~TeV at the Tevatron. This figure makes manifest the incompleteness of the description of collider bounds on contact operators in terms of a single minimum-scale constraint.

\section{Conclusions}
\label{sec:conc}

We have presented a detailed study of SMEFT effects in dilepton production, for the first time treating the EFT expansion in a mathematically consistent way by truncating the signal predictions at order $1/\Lambda^2$ and treating terms at order $1/\Lambda^4$ as theory uncertainties. The obtained bounds are notably weaker than those which arise from treatments of the SMEFT contributions to this process in which the $1/\Lambda^4$ terms originating from squared dimension-six amplitudes are kept and theory uncertainties are neglected, but they are robust representations of what can be said in a fully model-independent framework about the effect of generic NP on quark-lepton scatterings. We note in particular that these bounds only constrain two distinct linear combinations of operators, and are far from foreclosing the possibility of new contributions which cancel one another in this observable. In order to more fully constrain these operators, additional observables which are affected by different combinations of operators will need to be similarly constrained, allowing for a global fit of the various observables.

In this article we have neglected the effects of SM parameter shifts which arise due to another host of operators in the SMEFT, as they are subdominant effects at high energies. These have already been studied in detail in electroweak precision data, but just how much hadron collider data can contribute to that subset of operators, including theoretical errors on the SMEFT predictions consistently, will be addressed in a future article.

We find that the current LHC data, due to an underfluctuation of the data compared to the SM background rate, constrains primarily parameter points which further increase the predicted rate. The constraints from this are unexpectedly strong in some directions due to that fluctuation, and weaker in others. Presuming a reversion to the SM prediction in future data, we find that the ultimate LHC dataset of 3 ab$^{-1}$ will be able to constrain the majority of perturbative parameter space in the two-dimensional space of these linear combinations. Meanwhile, at lower energy scales, the Tevatron has provided important coverage where the perturbative series in $\Lambda^{-1}$ has broken down at the LHC, belying the traditional one-number claimed constraints for contact interactions. Tevatron constraints are key to closing parameter regions with cutoffs below $\Lambda\sim5$ TeV. For sensible scale choices (e.g.~10 TeV at the LHC) the breakdown of the EFT expansion corresponds to non-perturbative coupling values, in keeping with the intuition of renormalizable theories.

\section*{Acknowledgments}
The work of SA was supported by the DFG Research Training Group Symmetry Breaking in Fundamental Interactions (GRK~1581) and the Cluster of Excellence Precision Physics, Fundamental Interactions and Structure of Matter (PRISMA-EXC 1098). MK gratefully acknowledges support by the Swiss National Science Foundation (SNF) under contract 200021-175940. The work of WS was supported in part by the Alexander von Humboldt Foundation, in the framework of the Sofja Kovalevskaja Award 2016, endowed by the German Federal Ministry of Education and Research. 

\bibliography{Sources}

\end{document}